\documentclass[pra, twocolumn, showpacs, showkeys]{revtex4}

\usepackage{amsmath}
\usepackage{amssymb}
\usepackage[dvips]{graphicx}

\DeclareGraphicsExtensions{.ps}

\newcommand*{\etal}{\textit{et al.\ }}

\begin{document}
\title{Detuning effects in the one-photon mazer}
\date{\today}
\author{Thierry Bastin}
\email{T.Bastin@ulg.ac.be}
\author{John Martin}
\email{John.Martin@ulg.ac.be} \affiliation{Institut de Physique
Nucl\'eaire, Atomique et de Spectroscopie, Universit\'e de Li\`ege
au Sart Tilman, B\^at.\ B15, B - 4000 Li\`ege, Belgique}

\begin{abstract}
The quantum theory of the mazer in the non-resonant case (a
detuning between the cavity mode and the atomic transition
frequencies is present) is written. The generalization from the
resonant case is far from being direct. Interesting effects of the
mazer physics are pointed out. In particular, it is shown that the
cavity may slow down or speed up the atoms according to the sign
of the detuning and that the induced emission process may be
completely blocked by use of a positive detuning. It is also shown
that the detuning adds a potential step effect not present at
resonance and that the use of positive detunings defines a
well-controlled cooling mechanism. In the special case of a mesa
cavity mode function, generalized expressions for the reflection
and transmission coefficients have been obtained. The general
properties of the induced emission probability are finally
discussed in the hot, intermediate and cold atom regimes.
Comparison with the resonant case is given.
\end{abstract}



\maketitle

\section{Introduction}

The interaction of cold atoms with microwave high-$Q$ cavities (a
cold atom micromaser) has recently attracted increasing interest
since it was demonstrated by Scully \etal \cite{Scu96} that this
interaction leads to a new type of induced emission inside the
cavity. The new emission properties arise from the necessity to
treat quantum mechanically the center-of-mass motion of the atoms
interacting with the cavity. To insist on the importance of this
quantization, usually defined along the $z$ axis, the system was
called a mazer (for microwave amplification via $z$-motion-induced
emission of radiation). The complete quantum theory of the mazer
was described in a series of three papers by Scully and co-workers
\cite{Mey97, Lof97, Sch97}. The theory was written for two-level
atoms interacting with a single mode of the high-$Q$ cavity via a
one-photon transition. In particular, it was shown that the
induced emission properties are strongly dependent on the cavity
mode profile. Results were presented for the mesa, sech$^2$ and
sinusoidal modes. Retamal \etal \cite{Ret98} later refined these
results in the special case of the sinusoidal mode, and a
numerical method was proposed by Bastin and Solano \cite{Bas00}
for efficiently computing the mazer properties with arbitrary
cavity field modes. L\"offler \etal \cite{Lof98} showed also that
the mazer may be used as a velocity selection device for an atomic
beam. The mazer concept was extended by Zhang \etal \cite{Zha99,
Zha98, Zha99b}, who considered two-photon transitions
\cite{Zha99}, three-level atoms interacting with a single cavity
\cite{Zha98} and with two cavities \cite{Zha99b}. Collapse and
revival patterns with a mazer have been computed by Du \etal
\cite{Du99}. Arun \etal \cite{Aru00, Aru02} studied the mazer with
bimodal cavities and Agarwal and Arun \cite{Aga00} demonstrated
resonant tunneling of cold atoms through two mazer cavities.

In all these previous studies, the mazer properties were always
presented in the resonant case where the cavity mode frequency
$\omega$ is equal to the atomic transition frequency $\omega_0$.
When three-level atoms were considered, a generalized resonant
condition was assumed \cite{Zha98, Zha99b, Aru00}. In this paper,
we remove this restriction and establish the theory of the mazer
in the nonresonant case ($\omega \neq \omega_0$) for two-level
atoms.

The paper is organized as follows. In Sec.~\ref{ModelSection}, the
Hamiltonian modeling the mazer in the nonresonant case is
presented. The wave functions of the system are described and
generalized expressions for the reflection and transmission
coefficients in the special case of the mesa mode function are
derived. The properties of the induced emission probability when a
detuning is present are then discussed in Sec.~\ref{PemSection}.
Three regimes of the mazer are considered (hot, intermediate and
cold). A brief summary of our results is finally given in
Sec.~\ref{SummarySection}.

\section{Model}
\label{ModelSection}

\subsection{The Hamiltonian}

We consider a two-level atom moving along the $z$ direction on the
way to a cavity of length $L$. The atom is coupled unresonantly to
a single mode of the quantized field present in the cavity. The
atomic center-of-mass motion is described quantum mechanically and
the usual rotating-wave approximation is made. We thus consider
the Hamiltonian
\begin{equation}
    \label{Hamiltonian}
        H = \hbar \omega_0 \sigma^{\dagger} \sigma + \hbar \omega a^{\dagger} a + \frac{p^2}{2m}+ \hbar g \, u(z) (a^{\dagger} \sigma + a
        \sigma^{\dagger}),
\end{equation}
where $p$ is the atomic center-of-mass momentum along the $z$
axis, $m$ is the atomic mass, $\omega_0$ is the atomic transition
frequency, $\omega$ is the cavity field mode frequency, $\sigma =
|b \rangle \langle a|$ ($|a\rangle$ and $|b\rangle$ are,
respectively, the upper and lower levels of the two-level atom),
$a$ and $a^{\dagger}$ are, respectively, the annihilation and
creation operators of the cavity radiation field, $g$ is the
atom-field coupling strength, and $u(z)$ is the cavity field mode
function. We denote in the following the detuning $\omega -
\omega_0$ by $\delta$, the cavity field eigenstates by $|n\rangle$
and the global state of the atom-field system by
$|\psi(t)\rangle$.

\subsection{The wave functions}

We introduce the orthonormal basis
\begin{equation}
\label{basis} \begin{array}{l} |\Gamma_{n}^+(\theta)\rangle =
\cos\theta\,|a,n\rangle +
\sin\theta\,|b,n+1\rangle,\vspace{8pt}\\
|\Gamma_{n}^-(\theta)\rangle = -\sin\theta\,|a,n\rangle +
\cos\theta\,|b,n+1\rangle,
\end{array}
\end{equation}
with $\theta$ an arbitrary parameter. The
$|\Gamma_{n}^{\pm}(\theta)\rangle$ states coincide with the
noncoupled states $|a,n\rangle$ and $|b,n+1\rangle$ when $\theta =
0$ and with the dressed states when $\theta = \theta_n$ given by
\begin{equation}
    \label{thetads}
    \cot 2 \theta_n = - \frac{\delta}{\Omega_n}\, ,
\end{equation}
with
\begin{equation}
    \Omega_n = 2 g \sqrt{n + 1}.
\end{equation}

We denote as $|\pm, n\rangle$ the dressed states
$|\Gamma_{n}^{\pm}(\theta_n)\rangle$. The Schr\"odinger equation
reads in the $z$ representation and in the basis (\ref{basis})
\begin{widetext}
\begin{subequations}
\label{ss}
\begin{multline}\label{ss+}
i\hbar\frac{\partial}{\partial t}\psi^+_{n, \theta}(z,t)=\left[
  -\frac{\hbar^2}{2m}\frac{\partial^2}{\partial
    z^2}+(n+1)\hbar\omega-\cos^2\theta\:\hbar\delta\;+\;\hbar g
  u(z) \sqrt{n\!+\!1}\sin2\theta\right]\psi^+_{n, \theta}(z,t) \\
+\bigg[\hbar g u(z)
\sqrt{n\!+\!1}\;\cos2\theta+\frac{1}{2}\sin2\theta\:\hbar\delta\bigg]\psi^-_{n,
\theta}(z,t),
\end{multline}
\vspace{-0.5cm}
\begin{multline}\label{ss-}
i\hbar\frac{\partial}{\partial t}\psi^-_{n, \theta}(z,t)=\left[
  -\frac{\hbar^2}{2m}\frac{\partial^2}{\partial
    z^2}+(n+1)\hbar\omega-\sin^2\theta\:\hbar\delta\;-\;\hbar g
  u(z) \sqrt{n\!+\!1}\sin2\theta\right]\psi^-_{n, \theta}(z,t)  \\
+\bigg[\hbar g u(z)
\sqrt{n\!+\!1}\;\cos2\theta+\frac{1}{2}\sin2\theta\:\hbar\delta\bigg]\psi^+_{n,
\theta}(z,t),
\end{multline}
\end{subequations}
\end{widetext}
with
\begin{equation}
    \psi^{\pm}_{n, \theta}(z,t) = \langle z,
    \Gamma_{n}^{\pm}(\theta)| \psi(t)\rangle.
\end{equation}

We get for each $n$ two coupled partial differential equations. In
the resonant case ($\delta = 0$), these equations may be decoupled
over the entire $z$ axis when working in the dressed state basis
and the atom-field interaction reduces to an elementary scattering
problem over a potential barrier and a potential well defined by
the cavity (see Ref.~\cite{Mey97}). In the presence of detuning,
this is no longer the case~: there is no basis where
Eqs.~(\ref{ss}) would separate over the entire $z$ axis and the
interpretation of the atomic interaction with the cavity as a
scattering problem over two potentials is less evident.

Outside the cavity (which we define to be located in the range $0
< z < L$), the mode function $u(z)$ vanishes and Eqs.~(\ref{ss})
become in the noncoupled state basis ($\theta = 0$)
\begin{subequations}
\label{ssa}
\begin{eqnarray}
i\hbar\frac{\partial}{\partial t}\psi^a_n(z,t) & = & \left[
  -\frac{\hbar^2}{2m}\frac{\partial^2}{\partial
    z^2} \right]\psi^a_n(z,t),\\
i\hbar\frac{\partial}{\partial t}\psi^b_{n + 1}(z,t) & = & \left[
  -\frac{\hbar^2}{2m}\frac{\partial^2}{\partial
    z^2} + \hbar \delta \right]\psi^b_{n +
    1}(z,t),
\end{eqnarray}
\end{subequations}
with
\begin{subequations}
\label{psiapsib}
\begin{eqnarray}
\psi^a_n(z,t) & = & e^{i(\omega_0 + n \omega) t} \langle z,a,n|\psi(t)\rangle, \phantom{\bigg|}\\
\psi^b_{n + 1}(z,t) & = & e^{i(\omega_0 + n \omega) t} \langle z,
b,n+1|\psi(t)\rangle.
\end{eqnarray}
\end{subequations}

In Eqs.~(\ref{psiapsib}), we have introduced the exponential
factor $e^{i(\omega_0 + n \omega) t}$ in order to define the
energy scale origin at the $|a,n\rangle$ level. The solutions to
Eqs.~(\ref{ssa}) are obviously given by linear combinations of
plane wave functions. If we assume initially a monokinetic atom
(with momentum $\hbar k$) coming upon the cavity from the left
side (negative $z$ values) in the excited state $|a\rangle$ and
the cavity field in the number state $|n\rangle$, the atom-field
system is described outside the cavity by the wave function
components (which correspond to the eigenstate $|\phi_k\rangle$ of
energy $E_k = \hbar^2 k^2 / 2 m$)
\begin{subequations}
\label{psiapsibn}
\begin{eqnarray}
\psi^a_n(z,t) & = & e^{-i \frac{\hbar k^2}{2 m} t} \varphi^a_n(z), \phantom{\bigg|}\\
\psi^b_{n+1}(z,t) & = & e^{-i \frac{\hbar k^2}{2 m} t}
\varphi^b_{n+1}(z),
\end{eqnarray}
\end{subequations}
with
\begin{align}
&\varphi^a_n(z)\;\,=\;\,\left\{\begin{array}{ll}
e^{ikz}+\rho^a_n\,e^{-ikz}&\;z < 0,\vspace{0.2cm}\\
\tau^a_n \,e^{ik(z-L)}&\;z > L,
\end{array}\right.\\\nonumber\\
&\varphi^b_{n+1}(z)=\left\{\begin{array}{ll}
\rho^b_{n+1} \,e^{-ik_bz}&\;\;z < 0,\vspace{0.2cm}\\
\tau^b_{n+1} \,e^{ik_b(z-L)}&\;\;z > L,
\end{array}\right.
\end{align}
and
\begin{equation}
k^2_b=k^2-\frac{2m\delta}{\hbar}.
\end{equation}

Introducing
\begin{equation}
    \kappa^2 = \frac{2 m g}{\hbar}
\end{equation}
we may write
\begin{equation}\label{cab}
k^2_b=k^2-\kappa^2 \frac{\delta}{g}.
\end{equation}

\begin{figure}
\begin{center}
\noindent\mbox{\includegraphics[width=8cm, bb= 110 75 690 410,
clip = true]{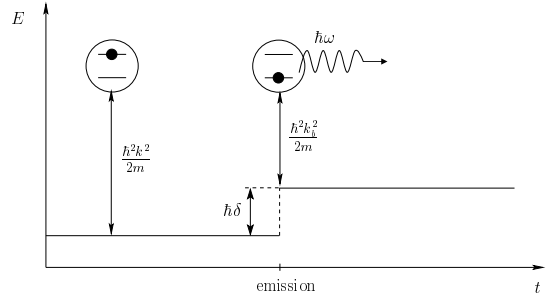}}
\end{center}
\vspace{-0.6cm} \caption{Potential step effect of the cavity when
a photon is emitted by the atom. $E$ represents the total energy
of the atom-field system.} \label{step}
\end{figure}

The solutions (\ref{psiapsibn}) must be interpreted as follows~:
the excited atom coming upon the cavity will be found reflected in
the upper state or in the lower state with the amplitude
$\rho^a_n$ and $\rho^b_{n+1}$, respectively, or transmitted with
the amplitude $\tau^a_n$ or $\tau^b_{n+1}$. However, in contrast
to the resonant case, the atom reflected or transmitted in the
lower state $|b\rangle$ will be found to propagate with a momentum
$\hbar k_b$ different from its initial value $\hbar k$. The atomic
transition $|a\rangle \rightarrow |b\rangle$ induced by the cavity
is responsible for a change of the atomic kinetic energy.
According to the sign of the detuning [see Eq.~(\ref{cab})], the
cavity will either speed up the atom (for $\delta < 0$) or slow it
down (for $\delta
> 0$). This results merely from energy conservation. When,
after leaving the cavity region, the atom is passed from the
excited state $|a\rangle$ to the lower state $|b\rangle$, the
photon number has increased by one unit in the cavity and the
internal energy of the atom-field system has varied by the
quantity $\hbar \omega - \hbar \omega_0 = \hbar \delta$. This
variation needs to be exactly counterbalanced by the external
energy of the system, i.e., the atomic kinetic energy. In this
sense, when a photon is emitted inside the cavity by the atom, the
cavity acts as a potential step $\hbar \delta$ (see
Fig.~\ref{step}), and the atom experiences an attractive or a
repulsive force according to the sign of the detuning. Similar
(although not identical) mechanical effects are obtained under the
adiabatic approximation \cite{Har91} (requiring no quantum
treatment of the atomic center-of-mass motion). In this case, the
dressed levels may also decelerate or accelerate the atoms.
However, in this regime and contrary to what is described here,
the atom always leaves the cavity with the same kinetic energy (if
no dissipation process is considered) and the mechanical effects
are not related to the emission of a photon inside the cavity.

Presently the use of positive detunings in the atom-field
interaction defines a well-controlled cooling mechanism. A single
excitation exchange between the atom and the field inside the
cavity is sufficient to cool the atom to a desired temperature $T
= \hbar^2 k_b^2/2 m k_B$ ($k_B$ is the Boltzmann constant) which
may in principle be as low as imaginable. However, if the initial
atomic kinetic energy $\hbar^2 k^2/2m$ is lower than $\hbar
\delta$ (i.e., if $k/\kappa < \sqrt{\delta/g}$), the transition
$|a,n\rangle \rightarrow |b,n+1\rangle$ cannot take place (as it
would remove $\hbar \delta$ from the kinetic energy) and no photon
can be emitted inside the cavity. In this case the emission
process is completely blocked.

Due to this change in the kinetic energy when a photon is emitted,
the reflection and transmission probabilities of the atom in the
lower state $|b\rangle$ are given, respectively, by (for $\hbar^2
k^2/2m
 > \hbar \delta$)
\begin{subequations}
\label{RbandTbn}
\begin{align}
    R^b_{n+1} & = \frac{k_b}{k}|\rho^b_{n+1}|^2, \\
    T^b_{n+1} & = \frac{k_b}{k}|\tau^b_{n+1}|^2.
\end{align}
\end{subequations}

These probabilities vanish for $\hbar^2 k^2/2m
 \leq \hbar \delta$. When the atom remains in the excited state $|a\rangle$ after
having interacted with the cavity, there is no change in the
atomic kinetic energy and the reflection and transmission
probabilities are directly given by
\begin{subequations}
\begin{align}
    R^a_n & = |\rho^a_n|^2, \\
    T^a_n & = |\tau^a_n|^2.
\end{align}
\end{subequations}

To calculate any quantity related to the reflection and
transmission coefficients, we must solve the Schr\"odinger
equation over the entire $z$ axis. Inside the cavity, the problem
is much more complex since we have two coupled partial
differential equations. In the special case of the mesa mode
function [$u(z) = 1$ inside the cavity, 0 elsewhere], the problem
is, however, greatly simplified. In the dressed state basis
($\theta = \theta_n$), the Schr\"odinger equations (\ref{ss}) take
the following form inside the cavity~:
\begin{equation}
\label{seqinside} i\hbar\frac{\partial}{\partial
t}\psi^{\pm}_n(z,t) = \left[
  -\frac{\hbar^2}{2m}\frac{\partial^2}{\partial
    z^2}+V^{\pm}_n\right]\psi^{\pm}_n(z,t)
\end{equation}
with
\begin{equation}
    \label{psipm}
    \psi^{\pm}_n(z,t) = e^{i(\omega_0 + n \omega) t} \langle
    z,\pm,n|\psi(t)\rangle
\end{equation}
and
\begin{subequations}
\begin{align}
    V^+_n & = \sin^2 \theta_n \, \hbar \delta + \hbar g \sqrt{n + 1} \, \sin 2 \theta_n,\phantom{\bigg|} \\
    V^-_n & = \hbar \delta - V^+_n. \label{VnmasVnp}
\end{align}
\end{subequations}

$V^+_n$ and $V^-_n$ represent the internal energies of the
$|+,n\rangle$ and $|-,n\rangle$ components, respectively. Except
for the resonant case, they cannot be strictly interpreted as a
potential barrier and a potential well as Eq.~(\ref{seqinside})
holds only inside the cavity.

Using Eq.~(\ref{thetads}), we have the well-known relations
\begin{equation}\label{cot}
\begin{array}{l}
\sin\theta_{n}=\frac{\sqrt{\Lambda_n+\delta}}{\sqrt{2\Lambda_n}},
\vspace{0.3cm}\\
\cos\theta_{n}=\frac{\sqrt{\Lambda_n-\delta}}{\sqrt{2\Lambda_n}},
\end{array}\quad
\begin{array}{l}
\tan\theta_{n}=\sqrt{\frac{\Lambda_{n}+\delta}{\Lambda_{n}-\delta}},\vspace{0.3cm}\\
\cot\theta_{n}=\sqrt{\frac{\Lambda_{n}-\delta}{\Lambda_{n}+\delta}},
\end{array}
\end{equation}
with
\begin{equation}
\Lambda_n=\sqrt{\delta^2+\Omega_n^2}.
\end{equation}

We thus have
\begin{subequations}
\begin{align}
    V^+_n & = \hbar g \sqrt{n + 1} \tan \theta_n, \phantom{\bigg|} \\
    V^-_n & = -\hbar g \sqrt{n + 1} \cot \theta_n.
\end{align}
\end{subequations}

\begin{figure}
\begin{center}
\noindent\mbox{\includegraphics[width=8cm, bb= 120 -35 670 750,
clip = true]{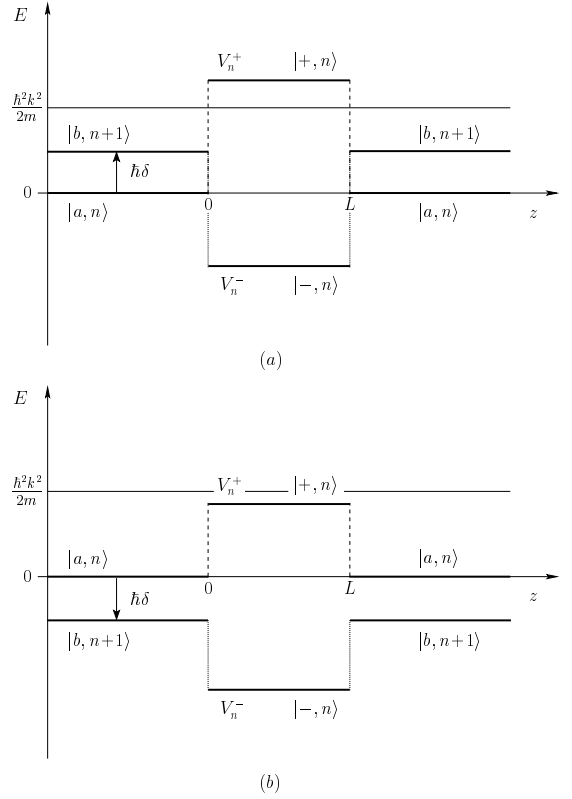}}
\end{center}
\vspace{-0.6cm} \caption{Schematic energy diagram of the
atom-field system inside and outside the cavity, for $\delta > 0$
(a) and $\delta < 0$ (b).} \label{intEnergies}
\end{figure}

The exponential factor $e^{i(\omega_0 + n \omega) t}$ has been
introduced as well in Eq.~(\ref{psipm}) in order to define the
same energy scale inside and outside the cavity. Figure
\ref{intEnergies} illustrates the internal energies of the
atom-field system over the whole $z$ axis. The positive internal
energy $V^+_n$ increases with positive detunings and vice versa
with negative ones. For a fixed value of the incident kinetic
energy $E_k = \hbar^2 k^2 / 2 m$, $V^+_n$ may be switched in this
way from a higher value than $E_k$ to a lower one. For large
positive (negative) detunings, $V^+_n$ tends to the
$|b,n+1\rangle$ ($|a,n\rangle$) state energy.

The most general solution of Eqs.~(\ref{seqinside}) corresponding
to the eigenstate $|\phi_k\rangle$ is given by
\begin{equation}
\label{psipminside}
\psi^{\pm}_n(z,t)= e^{-i \frac{\hbar k^2}{2 m}
t} \varphi^{\pm}_n(z)
\end{equation}
with
\begin{equation}
\varphi^{\pm}_n(z) = A^{\pm}_n e^{i k^{\pm}_n z} + B^{\pm}_n e^{-i
k^{\pm}_n z},
\end{equation}
where $A^{\pm}_n$ and $B^{\pm}_n$ are complex coefficients and
\begin{equation}
{k^{\pm}_n}^2 = k^2 - \frac{2m}{\hbar^2}V^{\pm}_n.
\end{equation}

Defining
\begin{equation}
    \kappa_n = \kappa \sqrt[4]{n + 1},
\end{equation}
we have
\begin{subequations}
\begin{align}
    {k^+_n}^2 & = k^2 - \kappa_n^2 \tan \theta_n, \phantom{\bigg|} \\
    {k^-_n}^2 & = k^2 + \kappa_n^2 \cot \theta_n.
\end{align}
\end{subequations}

Using Eq.~(\ref{VnmasVnp}), we may also write
\begin{equation}
    {k^-_n}^2 = k_b^2 + \kappa_n^2 \tan \theta_n.
\end{equation}

From Eq.~(\ref{basis}), we may express the wave function
components of the atom-field state inside the cavity over the
noncoupled state basis. We have
\begin{subequations}
\begin{align}
    \psi^a_n(z,t) & = \cos \theta_n \psi^+_n(z,t) - \sin \theta_n
    \psi^-_n(z,t), \phantom{\bigg|}\\
    \psi^b_{n+1}(z,t) & = \sin \theta_n \psi^+_n(z,t) + \cos \theta_n
    \psi^-_n(z,t).
\end{align}
\end{subequations}

This allows us to find the wave function components of the
eigenstate $|\phi_k\rangle$ over the entire $z$ axis. The
relations (\ref{psiapsibn}) hold with
\begin{align}\label{glob1}
&\varphi^a_n(z)\;\,=\;\,\left\{\begin{array}{ll}
e^{ikz}+\rho^a_n\, e^{-ikz}&\;z<0,\vspace{0.2cm}\\
\varphi^a_n(z)\big|_C&\;0<z<L,\vspace{0.2cm}\\
\tau^a_n \,e^{ik(z-L)}&\;z>L,
\end{array}\right.\\\nonumber\\
&\varphi^b_{n+1}(z)=\left\{\begin{array}{ll}\label{glob2}
\rho^b_{n+1}\, e^{-ik_bz}&\;\;z<0,\vspace{0.2cm}\\
\varphi^b_{n+1}(z)\big|_C&\;\;0<z<L,\vspace{0.2cm}\\
\tau^b_{n+1}\, e^{ik_b(z-L)}&\;\;z>L,
\end{array}\right.
\end{align}
and
\begin{subequations}
\label{phianzphibnz}
\begin{align}
\varphi^a_n(z)\big|_C & =\cos\theta_{n}\left(A^+_n e^{ik^+_n
z}+B^+_n e^{-ik^+_n z}\right) \nonumber \\ & \quad
-\sin\theta_{n}\left(A^-_n e^{ik^-_n z}+B^-_n e^{-ik^-_n
z}\right),\phantom{\bigg|}
\\
\varphi^b_{n+1}(z)\big|_C & =\sin\theta_{n}\left(A^+_n e^{ik^+_n
z}+B^+_n e^{-ik^+_n z}\right) \nonumber \\ & \quad
+\cos\theta_{n}\left(A^-_n e^{ik^-_n z}+B^-_n e^{-ik^-_n
z}\right).
\end{align}
\end{subequations}

The coefficients $\rho^a_n$, $\rho^b_{n+1}$, $\tau^a_n$,
$\tau^b_{n+1}$, $A^+_n$, $A^-_n$, $B^+_n$, and $B^-_n$ in
expressions (\ref{glob1})-(\ref{phianzphibnz}) are found by
imposing the continuity conditions on the wave function and its
first derivative at the cavity interfaces. A tedious calculation
yields
\begin{widetext}
\begin{align}
    \tau^a_n & = \frac{\cos^2 \theta_n \frac{\tau^-_n(k)}{\tau^-_n(k_b)} \, \tau^+_n(k_b) + \sin^2 \theta_n \, {\tau^-_n(k)}}{\left( \cos^2 \theta_n \frac{k - k_b}{k^c_n(L)} -1 \right) \left( \cos^2 \theta_n \frac{k - k_b}{k^t_n(L)} -1 \right)}, \\
    \rho^a_n & = \frac{\cos^2 \theta_n \frac{\tau^-_n(k)}{\tau'^-_n(k,k_b)}\frac{\Delta^+_n(k)}{\Delta^+_n(k_b)} \rho^+_n(k_b) + \left(1 - \cos^2 \theta_n \frac{\tau^+_n(k_b)}{\tau''^+_n(k,k_b)} \right) \rho^-_n(k) + \frac{\cos^2 \theta_n}{4}\left( \frac{k_b}{k} - \frac{k}{k_b} \right) u_n(k) \frac{\rho^-_n(k) \rho^+_n(k_b)}{\Delta^-_n(k) \Delta^+_n(k_b)} }{\left( \cos^2 \theta_n \frac{k -
k_b}{k^c_n(L)} -1 \right) \left( \cos^2 \theta_n \frac{k -
k_b}{k^t_n(L)} -1
    \right)},\\
    \tau^b_{n + 1} & = \frac{\sin 2 \theta_n}{4} \left( 1 + \frac{k}{k_b}
    \right)\frac{\frac{\tau^-_n(k)}{\tilde{\tau}^-_n(k,k_b)} \, \tau^+_n(k_b) - \frac{\tau^+_n(k_b)}{\tilde{\tau}^+_n(k,k_b)} \, \tau^-_n(k)}{\left( \cos^2 \theta_n \frac{k - k_b}{k^c_n(L)} -1 \right) \left( \cos^2 \theta_n \frac{k - k_b}{k^t_n(L)} -1
    \right)},\\
    \rho^b_{n + 1} & =\sin 2 \theta_n \frac{\frac{1}{2} \frac{\tau^-_n(k)}{\bar{\tau}^-_n(k, k_b)} \frac{\Delta^+_n(k)}{\Delta^+_n(k_b)} \rho^+_n(k_b) - \frac{1}{2} \frac{\tau^+_n(k_b)}{\bar{\tau}^+_n(k,k_b)} \rho^-_n(k) + \frac{1}{4} \left( \frac{k}{k_b} - 1 \right) v_n(k) \tau^-_n(k) \tau^+_n(k_b) }{\left( \cos^2 \theta_n \frac{k - k_b}{k^c_n(L)} -1 \right) \left( \cos^2 \theta_n \frac{k - k_b}{k^t_n(L)} -1
    \right)},
\end{align}
with
\begin{align}
u_n(k) & = \sin^2 \theta_n \frac{S^{+-}_n}{\sin(k^+_n L) \sin(k^-_n L)} + \left( \frac{k^-_n k^+_n}{k^2} - \frac{k^2}{k^-_n k^+_n} \right), \\
v_n(k) & = i \left( \frac{k^+_n}{k} \sin(k^+_n L) \cos(k^-_n L) - \frac{k^-_n}{k} \sin(k^-_n L) \cos(k^+_n L) \right) - \frac{\cos 2 \theta_n}{2} S^{+-}_n, \\
S^{+-}_n & = \sin(k^+_n L) \sin(k^-_n L) \left(\frac{k^-_n}{k^+_n} + \frac{k^+_n}{k^-_n} \right) + 2(\cos(k^-_n L) \cos(k^+_n L) - 1),
\end{align}
\end{widetext}
and
\begin{equation}
\rho^{\pm}_n(k)=i \Delta^{\pm}_n(k) \sin(k^{\pm}_n L)
\tau^{\pm}_n(k),
\end{equation}
\begin{equation}
\tau^{\pm}_n(k)=\left[ \cos(k^{\pm}_n L) - i \Sigma^{\pm}_n(k)
\sin(k^{\pm}_n L) \right]^{-1},
\end{equation}
\begin{equation}
\Sigma^{\pm}_n(k) = \frac{1}{2}\left( \frac{k^{\pm}_n}{k} +
\frac{k}{k^{\pm}_n} \right),
\end{equation}
\begin{equation}
\Delta^{\pm}_n(k) = \frac{1}{2}\left( \frac{k^{\pm}_n}{k} -
\frac{k}{k^{\pm}_n} \right),
\end{equation}
\begin{equation}
\tau'^{\pm}_n(k,k_b)=\left[ \cos(k^{\pm}_n L) - i \frac{k_b}{k}
\Sigma^{\pm}_n(k) \sin(k^{\pm}_n L) \right]^{-1}\!\!\!\!,
\end{equation}
\begin{equation}
\tau''^{\pm}_n(k,k_b)=\left[ \cos(k^{\pm}_n L) - i \frac{k}{k_b}
\Sigma^{\pm}_n(k) \sin(k^{\pm}_n L) \right]^{-1}\!\!\!\!,
\end{equation}
\begin{equation}
\tilde{\tau}^{\pm}_n(k,k_b)=\left[ \cos(k^{\pm}_n L) - i
\tilde{\Sigma}^{\pm}_n(k, k_b) \sin(k^{\pm}_n L) \right]^{-1},
\end{equation}
\begin{equation}
\bar{\tau}^{\pm}_n(k,k_b)=\left[ \cos(k^{\pm}_n L) - i \frac{k +
k_b}{2 k_b} \Sigma^{\pm}_n(k) \sin(k^{\pm}_n L)
\right]^{-1}\!\!\!\!,
\end{equation}
\begin{equation}
\tilde{\Sigma}^{\pm}_n(k, k_b) = \left( \frac{k^{\pm}_n}{k + k_b}
+ \frac{k_b}{k + k_b} \frac{k}{k^{\pm}_n} \right),
\end{equation}
\begin{equation}
k^c_n(L) = i \frac{\left(k + i \cot(\frac{k^-_n L}{2}) k^-_n
\right) \left( k_b + i \cot(\frac{k^+_n L}{2}) k^+_n\right)}{
\cot(\frac{k^-_n L}{2}) k^-_n - \cot(\frac{k^+_n L}{2}) k^+_n},
\end{equation}
\begin{equation}
k^t_n(L) = i \frac{\left(k - i \tan(\frac{k^-_n L}{2}) k^-_n
\right) \left( k_b - i \tan(\frac{k^+_n L}{2})
k^+_n\right)}{\tan(\frac{k^+_n L}{2}) k^+_n - \tan(\frac{k^-_n
L}{2}) k^-_n}.
\end{equation}

At resonance, $\delta = 0$, $\theta_n = \pi/4$, $k_b = k$,
$\tau^{\pm}_n(k) = \tau'^{\pm}_n(k, k_b) = \tau''^{\pm}_n(k, k_b)
= \tilde{\tau}^{\pm}_n(k, k_b) = \bar{\tau}^{\pm}_n(k, k_b)$ and
the reflection and transmission coefficients reduce to the
well-known results (see Ref.~\cite{Mey97})
\begin{subequations}
\begin{align}
\tau^a_n = \frac{1}{2} \left( \tau^+_n + \tau^-_n \right), \\
\rho^a_n = \frac{1}{2} \left( \rho^+_n + \rho^-_n \right),
\end{align}
\end{subequations}
and
\begin{subequations}
\begin{align}
\tau^b_{n + 1} = \frac{1}{2} \left( \tau^+_n - \tau^-_n \right), \\
\rho^b_{n + 1} = \frac{1}{2} \left( \rho^+_n - \rho^-_n \right).
\end{align}
\end{subequations}

\section{Induced emission probability}
\label{PemSection}

The induced emission probability of a photon inside the cavity is
given by
\begin{equation}
    \mathcal{P}_{\textrm{em}}(n) = R^b_{n+1} + T^b_{n+1}.
\end{equation}

According to Eqs.~(\ref{RbandTbn}), this probability may be
written as
\begin{equation}\label{Pemgen}
\mathcal{P}_{\textrm{em}}(n)=\left\{
\begin{array}{ll}
\frac{k_b}{k}\Big(\,|\rho^b_{n+1}|^2+|\tau^b_{n+1}|^2\,\Big) &
\text{if }\frac{k}{\kappa}>\sqrt{\frac{\delta}{g}},\vspace{0.2cm}\\
0 & \text{otherwise.}
\end{array}\right.
\end{equation}

We distinguish three regimes determined by the incident kinetic
energy of the atom compared with the internal energy $V_n^+$ (see
Fig.~\ref{intEnergies})~: the hot atom regime (when $k \gg
\kappa_n \sqrt{\tan \theta_n}$ and $k_b$ is approximated to $k$),
the intermediate regime ($k \approx \kappa_n \sqrt{\tan
\theta_n}$) and the cold atom regime ($k \ll \kappa_n \sqrt{\tan
\theta_n}$). In comparison with the resonant case, the detuning
defines an additional parameter that fixes the working regime of
the system.

\subsection{Hot atom regime}

In the hot atom regime, the kinetic energy is much higher than the
energies $V_n^{\pm}$ and the atoms are always transmitted through
the cavity. For the mesa mode, the atomic momentum inside the
cavity is given by
\begin{subequations}
\begin{align}
\hbar k_n^+ \simeq \hbar k \left( 1 - \frac{\kappa_n^2}{2 k^2}
\tan \theta_n \right), \\
\hbar k_n^- \simeq \hbar k \left( 1 + \frac{\kappa_n^2}{2 k^2}
\cot \theta_n \right).
\end{align}
\end{subequations}

After the atom-field interaction, the global state of the system
reduces to
\begin{equation}
\begin{split}
    |\psi(t)\rangle = \int \! dz \, \psi(z,t) & \left( \cos \theta_n e^{-i \frac{\kappa_n^2 L}{2 k} \tan \theta_n}|z,+,n\rangle \right. \\
                      & \left. -\sin \theta_n e^{i \frac{\kappa_n^2 L}{2 k} \cot \theta_n}|z,-,n\rangle \right)
\end{split}
\end{equation}
with
\begin{equation}
    \psi(z,t) = e^{i k z} e^{-i \frac{\hbar k^2}{2 m}t}.
\end{equation}

The induced emission probability is thus given by
\begin{equation}
\begin{split}
    \mathcal{P}_{\textrm{em}}(n) & = |\langle z,b,n+1|\psi(t)\rangle|^2 \\
    & = \frac{\sin^2(2 \theta_n)}{4} \left| \, e^{-i \frac{\kappa_n^2 L}{2 k} \tan \theta_n} - e^{i \frac{\kappa_n^2 L}{2 k} \cot
    \theta_n}\right|^2.
\end{split}
\end{equation}

Using Eq.~(\ref{cot}) we get straightforwardly
\begin{equation}
    \label{PemRabi}
    \mathcal{P}_{\textrm{em}}(n) = \left[ 1 +
    \left(\frac{\delta}{\Omega_n}\right)^2\right]^{-1}\sin^2
    \left( \frac{\tau}{2} \sqrt{\Omega_n^2 + \delta^2}\right),
\end{equation}
where $\tau = \frac{m L}{\hbar k}$ is the classical transit time
of the thermal atoms through the cavity.

Equation (\ref{PemRabi}) is exactly the classical expression of
the induced emission probability of an atom interacting with a
single mode during a time $\tau$. We recover the well-known Rabi
oscillations in the general case where the field and the atomic
frequencies are detuned by the quantity $\delta$.

\subsection{Intermediate regime}

\begin{figure}
\begin{center}
\noindent\mbox{\includegraphics[width=8cm, bb= 118 490 490 724,
clip = true]{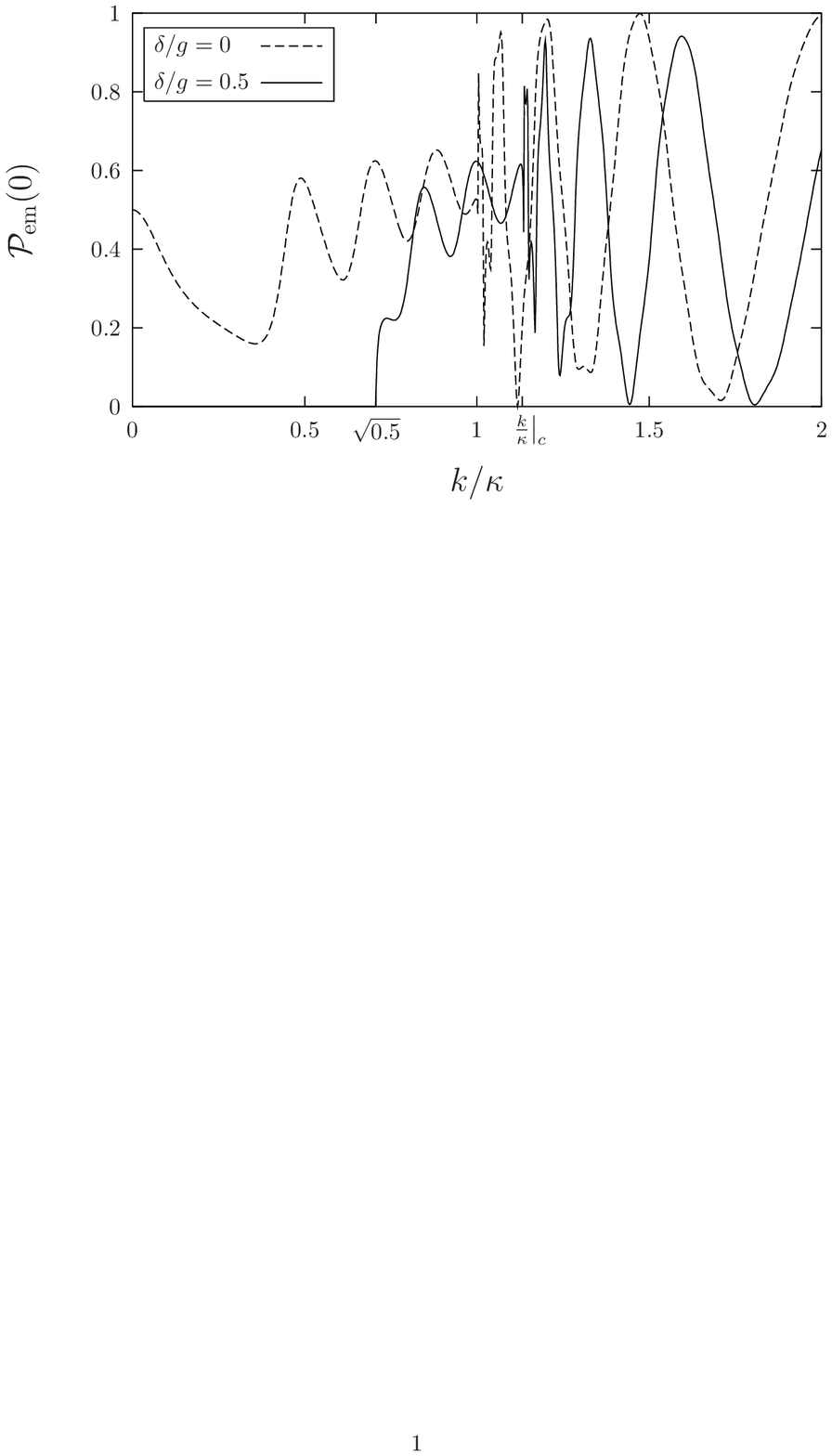}}
\end{center}
\vspace{-0.6cm} \caption{Induced emission probability
$\mathcal{P}_{\textrm{em}}(n=0)$ with respect to $k/\kappa$ (for
$\kappa L = 10 \pi$ and two different values of the detuning).}
\label{pemksk}
\end{figure}

\begin{figure}
\begin{center}
\noindent\mbox{\includegraphics[width=8cm, bb= 118 490 490 724,
clip = true]{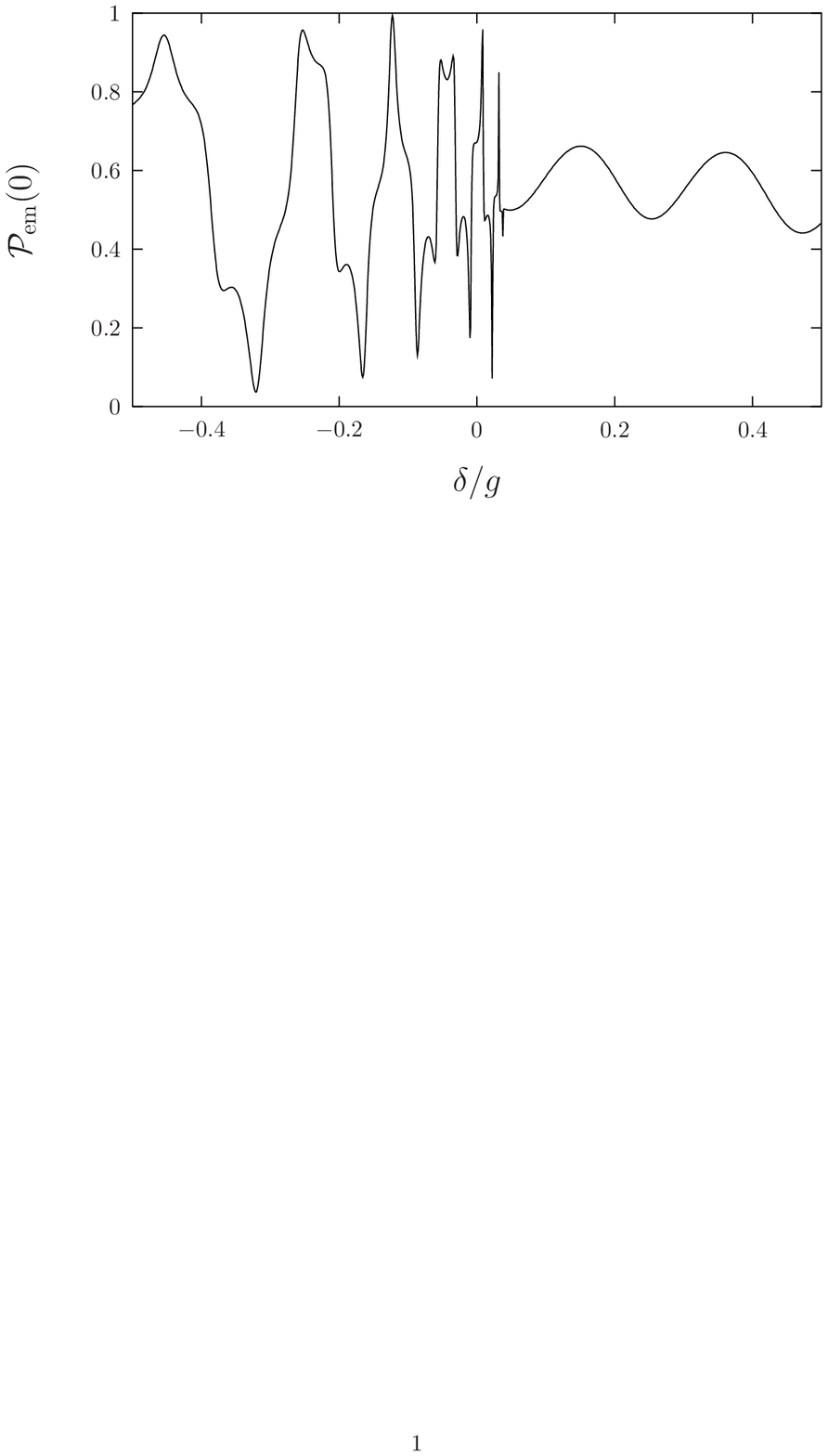}}
\end{center}
\vspace{-0.6cm} \caption{Induced emission probability
$\mathcal{P}_{\textrm{em}}(n=0)$ with respect to $\delta/g$ in the
intermediate regime ($k/\kappa = 1.01$). The interaction length
was fixed to $\kappa L = 100$.} \label{pemdsg}
\end{figure}

The frontier between the cold atom and the hot atom regime appears
when the atomic kinetic energy becomes equal to the positive
energy $V^+_n$ inside the cavity, i.e., when the ratio
$k/\kappa_n$ is equal to the critical value
\begin{equation}
    \label{criticalkonkappa}
    \left. \frac{k}{\kappa_n} \right|_c = \sqrt{\tan \theta_n} =\left(\frac{\sqrt{\left(\frac{\delta}{g}\right)^2 + 4 (n + 1)}\,+\,\frac{\delta}{g}}
    {\sqrt{\left(\frac{\delta}{g}\right)^2 + 4 (n + 1)}\,-\,\frac{\delta}{g}}\right)^{\frac{1}{4}}
\end{equation}

At resonance, this frontier occurs for $k/\kappa_n = 1$. This
condition changes significantly when the cavity and the atomic
transition are detuned. This is well illustrated in
Fig.~\ref{pemksk}, which shows the induced emission probability as
a function of $k/\kappa$ at resonance and for a positive value of
the detuning. In the second case, the regime change occurs for a
larger value of $k/\kappa$ than one. Please notice also that the
induced emission probability vanishes for $k/\kappa <
\sqrt{\delta/g}$ according to Eq.~(\ref{Pemgen}).

Equation (\ref{criticalkonkappa}) may be inverted to yield a
critical detuning when working with a fixed value of $k/\kappa$.
We get
\begin{equation}
\label{criticaldeltaong} \left. \frac{\delta}{g} \right|_c=
\sqrt{n + 1} \left[ \left(\frac{k}{\kappa_n}\right)^2 -
\left(\frac{k}{\kappa_n}\right)^{\!\!-2}\right].
\end{equation}

Figure \ref{pemdsg} illustrates the induced emission probability
as a function of the detuning for $k/\kappa$ slightly greater than
one. At resonance, the system is on the hot atom regime side as
the atomic kinetic energy is greater than the internal energy
$V^+_n$. When the detuning is increased, $V^+_n$ is increased as
well and becomes greater than the kinetic energy [see
Fig.~\ref{intEnergies}(a)], switching the system toward the cold
atom regime. This therefore could define a convenient way to
switch from one regime to the other rather than varying the
incident atomic momentum.

\begin{figure}
\begin{center}
\noindent\mbox{\includegraphics[width=8cm, bb= 118 490 490 724,
clip = true]{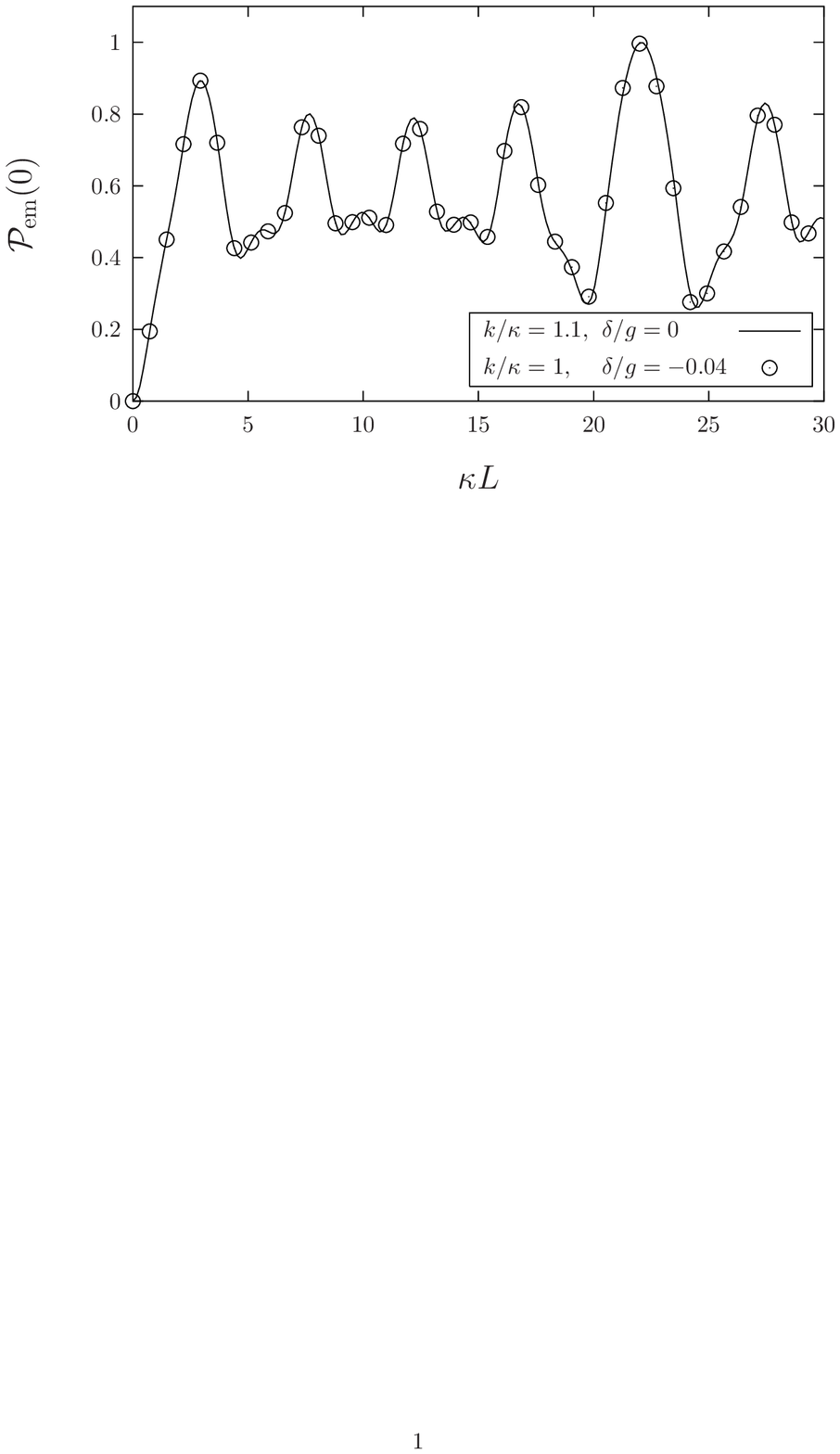}}
\end{center}
\vspace{-0.6cm} \caption{Induced emission probability
$\mathcal{P}_{\textrm{em}}(n=0)$ with respect to the interaction
length $\kappa L$ in the intermediate regime. Comparison of a
detuning variation with a change of the atomic kinetic energy.}
\label{peminter}
\end{figure}

A similar effect of the detuning is presented in
Fig.~\ref{peminter} which presents the induced emission
probability with respect to the interaction length in the
intermediate regime. It appears clearly there that working with a
negative detuning is similar to working with hotter atoms at
resonance. This results merely from the level of $V^+_n$ compared
to the incident kinetic energy, which is identical in both cases
considered in Fig.~\ref{peminter}.

From all these situations, we conclude that a detuning variation
has an identical effect than a change of the kinetic energy of the
incoming atoms, confirming that the only important parameter of
the system in the intermediate regime is the actual value of the
$V^+_n$ energy in comparison with the kinetic energy $\hbar^2
k^2/2 m$.

\subsection{Cold atom regime}

\begin{figure}
\begin{center}
\noindent\mbox{\includegraphics[width=8cm, bb= 118 490 490 724,
clip = true]{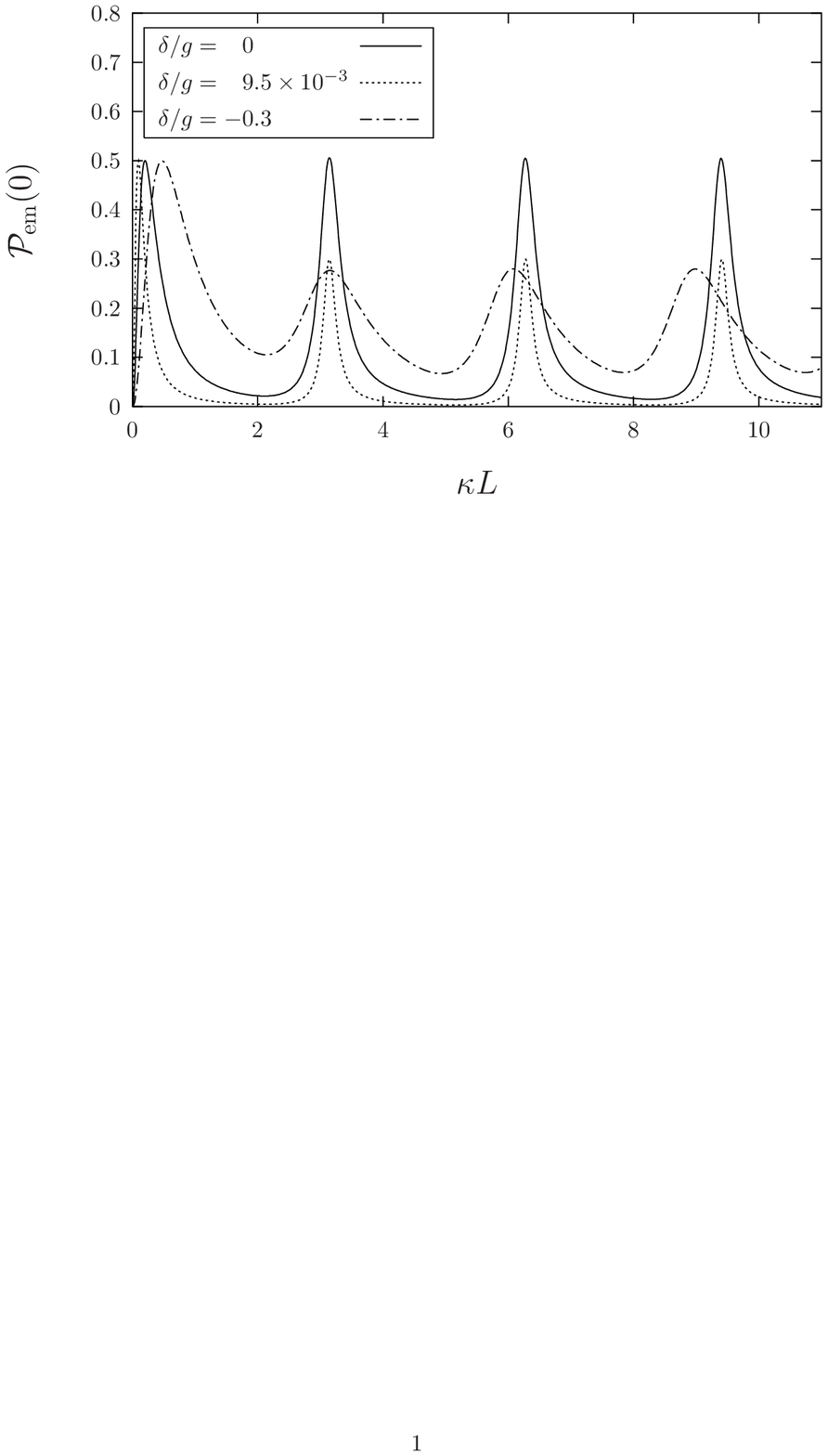}}
\end{center}
\vspace{-0.6cm} \caption{Induced emission probability
$\mathcal{P}_{\textrm{em}}(n=0)$ with respect to the interaction
length $\kappa L$ in the cold atom regime ($k/\kappa = 0.1$) and
for various values of the detuning.} \label{pemmazer}
\end{figure}

In the cold atom regime ($k/\kappa_n \ll \sqrt{\tan \theta_n}$),
the induced emission probability exhibits a completely different
behavior. We have in this regime for $|\delta|/\Omega_n \ll 1$,
$\exp(\kappa_n L) \gg 1$, and $\kappa_n L \ll (\kappa_n/k)^2$,
\begin{equation}
    \label{Pemcold}
    \mathcal{P}_{\textrm{em}}(n) = \frac{\mathcal{B}(L)}{2} \frac{1 + \frac{\cot \theta_n}{2} \sin(2 \kappa_n \sqrt{\cot \theta_n} L)}{1 + \left( \frac{\kappa_n}{2 k} \right)^2 \cot \theta_n \sin^2(\kappa_n \sqrt{\cot \theta_n}
    L)},
\end{equation}
with
\begin{equation}
    \mathcal{B}(L) = \frac{k_b}{k}\frac{1}{\left| \cos^2 \theta_n \frac{k - k_b}{k^c_n(L)} - 1 \right|^2 \left| \cos^2 \theta_n \frac{k - k_b}{k^t_n(L)} - 1
    \right|^2}.
\end{equation}

At resonance, this expression simplifies to
\begin{equation}
    \label{Pemcoldatresonance}
    \mathcal{P}_{\textrm{em}}(n) = \frac{1}{2} \frac{1 + \frac{1}{2} \sin(2 \kappa_n L)}{1 + \left( \frac{\kappa_n}{2 k} \right)^2 \sin^2(\kappa_n L)}
\end{equation}
and the results of Meyer \etal \cite{Mey97} are well recovered.
Figure \ref{pemmazer} illustrates the induced emission
probability~(\ref{Pemcold}) with respect to the interaction length
$\kappa L$ for various values of the detuning. The curves present
a series of peaks where the induced emission probability is
optimum. The detuning affects the peak position, amplitude, and
width. Similarly to the resonant case, the curves still look like
the Airy function of classical optics $[1 + F \sin^2(\Delta /
2)]^{-1}$ with finesse $F$ and total phase difference $\Delta$,
even if the structure of Eq.~(\ref{Pemcold}) has become
complicated with the factor $\mathcal{B}(L)$. In fact, this
equation is extremely well fitted in its domain of validity by the
function
\begin{equation}
    \label{PemcoldSimplified}
    \mathcal{P}_{\textrm{em}}(n) = \frac{2 k_b/k}{(1 + k_b/k)^2}\frac{1 + \frac{1}{2} \sin (2 \kappa_n \sqrt{\cot \theta_n} L)}{1 + (\frac{\kappa_n}{k_b + k})^2 \sin^2(\kappa_n \sqrt{\cot \theta_n}
    L)}.
\end{equation}

\subsubsection {Peak position}

The induced emission probability is optimum when
\begin{equation}
    \label{peakposition}
    \kappa_n \sqrt{\cot \theta_n} L = m \pi \quad (m
    \textrm{ a positive integer}).
\end{equation}

This occurs when the cavity length fits a multiple of one-half the
de Broglie wavelength $\lambda_{\mathrm{dB}}$ of the atom inside
the cavity~:
\begin{equation}
    \label{LeqldB}
    L = m \frac{\lambda_{\mathrm{dB}}}{2}.
\end{equation}

Indeed, in the cold atom regime, only the $|-,n\rangle$ component
propagates inside the cavity with the de Broglie wavelength
\begin{equation}
\label{lambdadB}
\lambda_{\mathrm{dB}}=\frac{2\pi}{k^-_{n}}\simeq
\frac{2\pi}{\kappa_n \sqrt{\cot\theta_{n}}}.
\end{equation}

Inserting Eq.~(\ref{lambdadB}) into Eq.~(\ref{LeqldB}) gives the
condition (\ref{peakposition}).

\subsubsection {Peak amplitude}

\begin{figure}
\begin{center}
\noindent\mbox{\includegraphics[width=8cm, bb= 118 490 490 724,
clip = true]{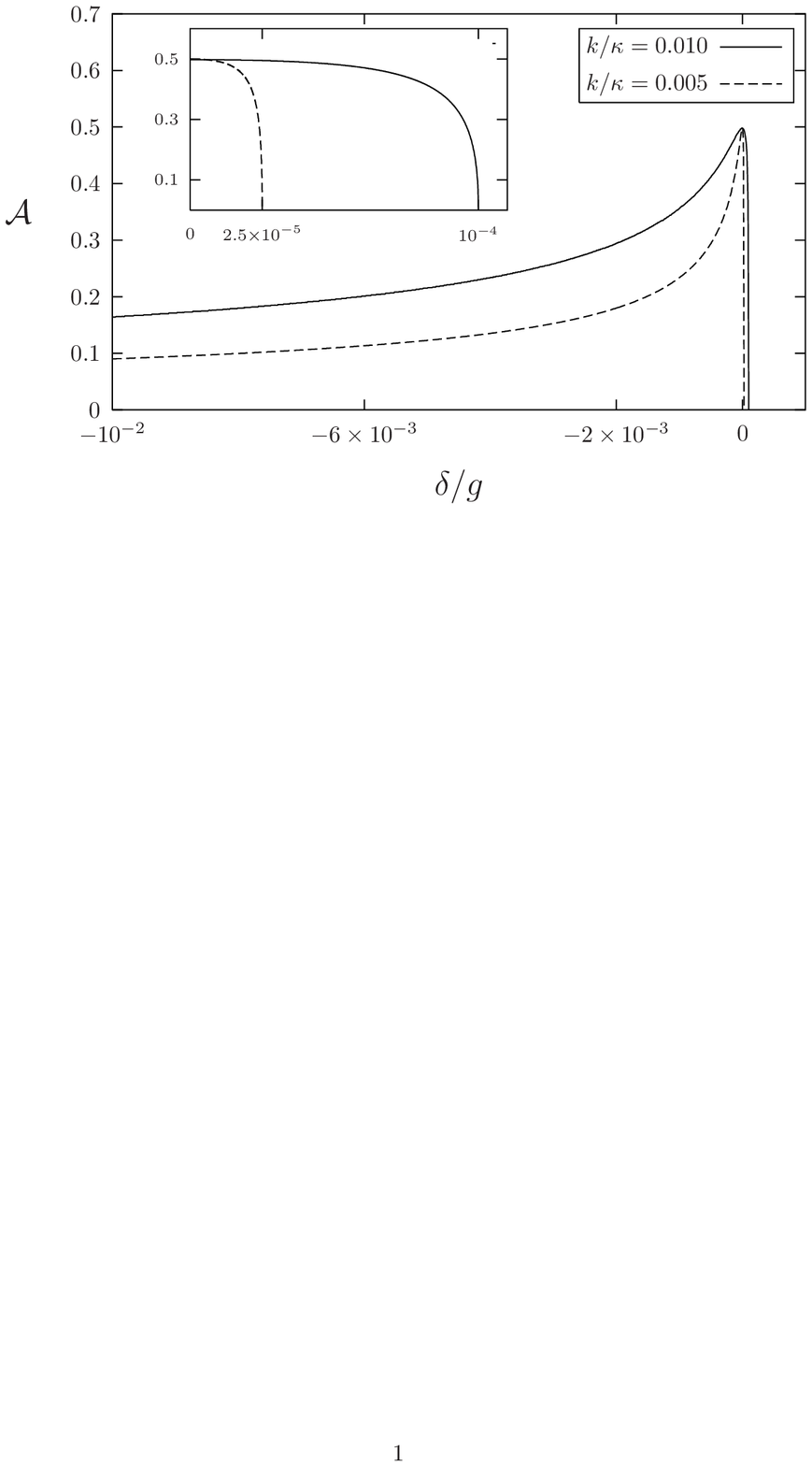}}
\end{center}
\vspace{-0.6cm} \caption{Amplitude $\mathcal{A}$ of the resonances
with respect to $\delta/g$ for two values of $k/\kappa$ in the
cold atom regime.} \label{PeakAmpFig}
\end{figure}

The peak amplitude of the induced emission probability
$\mathcal{P}_{\textrm{em}}(n)$ is given by
\begin{equation}
    \label{A}
    \mathcal{A} \equiv \frac{\mathcal{B}(L=m \frac{\lambda_{\mathrm{dB}}}{2})}{2}
    \simeq \frac{1}{2} \frac{4 k_b/k}{(1 + k_b/k)^2}.
\end{equation}

We illustrate this amplitude in Fig.~\ref{PeakAmpFig} as a
function of the detuning $\delta$. In contrast to the hot atom
regime [see Eq.~(\ref{PemRabi})], the curves present a strong
asymmetry with respect to the sign of the detuning. This results
from the potential step $\hbar \delta$ (see Sec.\ II.) experienced
by the atoms when they emit a photon. For cold atoms whose energy
is similar or less than the step height, the sign of the step is a
crucial parameter. According to Eq.~(\ref{Pemgen}), the induced
emission probability drops down very rapidly to zero for positive
detunings, in contrast to what happens for negative detunings.

It is also interesting to note that the peak amplitude (\ref{A})
is equal to the amplitude at resonance ($1/2$) times the factor
$(4 k_b/k)/(1 + k_b/k)^2$ that corresponds exactly to the
transmission factor of a particle of momentum $\hbar k$ through a
potential step $\hbar \delta$. This is an additional argument to
say that the use of a detuning adds a potential step effect for
the atoms emitting a photon inside the cavity (see
Fig.~\ref{step}).

\subsubsection{Peak width}

The peak width is determined by the finesse $(\frac{\kappa_n}{k_b
+ k})^2$. Positive detunings increase the finesse ($k_b < k$)
while negative ones decrease it ($k_b > k$).

\subsubsection{Large detunings}

\begin{figure}
\begin{center}
\noindent\mbox{\includegraphics[width=8cm, bb= 170 170 800 650,
clip = true]{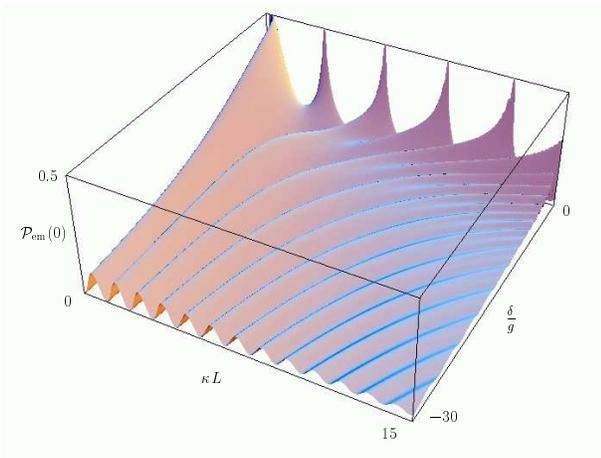}}
\end{center}
\vspace{-0.6cm} \caption{Induced emission probability
$\mathcal{P}_{\textrm{em}}(n=0)$ with respect to the interaction
length $\kappa L$ and $\delta/g$ (for $k/\kappa = 0.1$).}
\label{twoDPlot}
\end{figure}

For large detunings, Eq.~(\ref{Pemcold}) is no longer valid and
the induced emission probability must be computed using the
general relation~(\ref{Pemgen}). We present in Fig.~\ref{twoDPlot}
$\mathcal{P}_{\textrm{em}}(0)$ as a function of the detuning and
the interaction length $\kappa L$. The variation of the resonance
positions with respect to the detuning is very clear in this
figure. It is interesting to note that one resonance out of two
disappears when increasing the detuning toward negative values.
Also, the induced emission probability does not decrease
monotonically with the detuning (especially for small interaction
lengths). This effect is strictly limited to the cold atom regime
as for hot atoms the Rabi oscillation amplitudes always decrease
when larger and larger detunings are used [see
Eq.~(\ref{PemRabi})].

The use of large negative detunings in the cold atom regime is not
limitless. As $-\delta$ increases, the internal energy $V_n^+$
decreases [see Fig.~\ref{intEnergies}(b)] and may finally become
lower than the incident kinetic energy. To keep the system in the
cold atom regime, we must have for $k/\kappa_n \ll 1$ [see
Eq.~(\ref{criticaldeltaong})]
\begin{equation}
    -\frac{\delta}{g} < \sqrt{n + 1} \left( \frac{\kappa_n}{k}
    \right)^2.
\end{equation}

This condition is well respected on Fig.~\ref{twoDPlot} as the
limiting lower value of $\delta/g$ to keep the system in the cold
atom regime is $-100$ for $k/\kappa = 0.1$.

The use of large positive detunings in the cold atom regime is not
possible as the induced emission probability vanishes for
\begin{equation}
    \frac{\delta}{g} \geq \left( \frac{k}{\kappa} \right)^2.
\end{equation}

\section{Summary}
\label{SummarySection}

In this paper we have presented the quantum theory of the mazer in
the nonresonant case. Interesting effects have been pointed out.
In particular, we have shown that the cavity may slow down or
speed up the atoms according to the sign of the detuning and that
the induced emission process may be completely blocked by use of a
positive detuning. We have also demonstrated that the detuning
adds a potential step effect not present at resonance. This
defines a well-controlled cooling mechanism for positive
detunings. In the special case of the mesa mode function,
generalized expressions for the reflection and transmission
coefficients have been obtained. The properties of the induced
emission probability in the presence of a detuning have been
discussed. In the cold atom regime, we have obtained a simplified
expression for this probability and have been able to describe the
detuning effects on the resonance amplitude, width, and position.
In contrast to the hot atom regime, we have shown that the mazer
properties are not symmetric with respect to the sign of the
detuning. In the intermediate regime, the use of detuning could be
a convenient way to switch from the hot atom regime to the cold
atom one.

\begin{acknowledgments}
This work has been supported by the Belgian Institut
Interuniversitaire des Sciences Nucl\'eaires (IISN). T.~B. wants
to thank Prof.\ Dr.\ H. Walther and Dr.\ E. Solano for the
hospitality at Max-Planck-Institut f\"ur Quantenoptik in Garching
(Germany).
\end{acknowledgments}

\end{document}